\documentclass[10pt,preprintnumbers, floatfix, letterpaper, onecolumn,aps,prd,epsfig,nofootinbib,longbibliography]{revtex4-1}

\usepackage{graphicx}
\usepackage[utf8]{inputenc}
\usepackage[T1]{fontenc}
\usepackage{physics}
\usepackage{latexsym}
\usepackage{epstopdf}
\usepackage{latexsym}
\usepackage{amssymb}
\usepackage{amsmath}
\usepackage{mathtools}
\usepackage{color}
\usepackage{mathrsfs}
\usepackage{xparse}
\usepackage{enumerate}
\usepackage{multirow,tabularx,boldline,array}
\usepackage{bbding}
\usepackage{enumitem}
\usepackage{pifont}
\usepackage[center]{subfigure}
\usepackage[
            pdfstartview=FitH,
            bookmarksnumbered=true,
            bookmarksopen=true,
            colorlinks,
            linkcolor=blue,
            anchorcolor=green,
            citecolor=blue
            ]{hyperref}

\begin{document}

  \renewcommand\arraystretch{2}
 \newcommand{\bq}{\begin{equation}}
 \newcommand{\eq}{\end{equation}}
 \newcommand{\nb}{\nonumber}
 \newcommand{\lb}{\label}
 \newcommand{\cb}{\color{blue}}
    \newcommand{\cc}{\color{cyan}}
        \newcommand{\cm}{\color{magent_a}}
\newcommand{\rc}{\rho^{\scriptscriptstyle{\mathrm{I}}}_c}
\newcommand{\rd}{\rho^{\scriptscriptstyle{\mathrm{II}}}_c}
\NewDocumentCommand{\evalat}{sO{\big}mm}{%
  \IfBooleanTF{#1}
   {\mleft. #3 \mright|_{#4}}
   {#3#2|_{#4}}%
}
\newcommand{\PRL}{Phys. Rev. Lett.}
\newcommand{\PL}{Phys. Lett.}
\newcommand{\PR}{Phys. Rev.}
\newcommand{\CQG}{Class. Quantum Grav.}

\title{Entanglement island and Page curve for one-sided charged black hole }

\author{Yun-Feng Qu$^{1}$}
\email{QuyunFeng@hunnu.edu.cn}
\affiliation{	$^1$ Department of Physics and Synergetic Innovation Center for Quantum Effects and Applications, Hunan Normal University, 36 Lushan Rd., Changsha, Hunan 410081, China\\
	$^2$ Institute of Interdisciplinary Studies, Hunan Normal University, 36 Lushan Rd., Changsha, Hunan 410081, China}
	
\author{Yi-Ling Lan$^{1}$}
\email{Yiling$\_$Lan@hunnu.edu.cn}
\affiliation{	$^1$ Department of Physics and Synergetic Innovation Center for Quantum Effects and Applications, Hunan Normal University, 36 Lushan Rd., Changsha, Hunan 410081, China\\
	$^2$ Institute of Interdisciplinary Studies, Hunan Normal University, 36 Lushan Rd., Changsha, Hunan 410081, China}

\author{Hongwei Yu$^{1,2}$~\footnote{Corresponding author}}
\email{hwyu@hunnu.edu.cn}
\affiliation{	$^1$ Department of Physics and Synergetic Innovation Center for Quantum Effects and Applications, Hunan Normal University, 36 Lushan Rd., Changsha, Hunan 410081, China\\
	$^2$ Institute of Interdisciplinary Studies, Hunan Normal University, 36 Lushan Rd., Changsha, Hunan 410081, China}

\author{Wen-Cong Gan}
	\email{ganwencong@jxnu.edu.cn}
	\affiliation{College of Physics and Communication Electronics, Jiangxi Normal University, Nanchang, 330022, China}

\author{Fu-Wen Shu}
	\email{shufuwen@ncu.edu.cn}
	\affiliation{Department of Physics, Nanchang University, Nanchang, 330031, China\\Center for Relativistic Astrophysics and High Energy Physics, Nanchang University, Nanchang, 330031, China}


\begin{abstract}
In this paper, we extend the method of calculating the entanglement entropy of Hawking radiation of black holes using the ``in'' vacuum state, which describes one-sided asymptotically flat neutral black hole formed by gravitational collapse, to dynamic charged black holes. We explore the influence of charge on the position of the boundary of island $\partial I$ and the Page time. Due to their distinct geometric structures, we discuss non-extremal and extremal charged black holes separately. In non-extremal cases, the emergence of island saves the bound of entropy at late times, and the entanglement entropy of Hawking radiation satisfies the Page curve. Moreover, we also find that the position of the boundary of island $\partial I$  depends on the position of the cutoff surface (observers), differing  from the behavior in eternal charged black holes. In extremal black holes, when the island exists, the entanglement entropy is approximately equal to the Bekenstein-Hawking entropy, while the entanglement entropy becomes ill-defined when island is absent. Our analysis underscores how different geometric configurations significantly influence the behavior of entropy.
 \end{abstract}

\maketitle
\tableofcontents
\section{Introduction}
 \renewcommand{\theequation}{1.\arabic{equation}}\setcounter{equation}{0}

The black hole information paradox \cite{Hawking:1976ra} is a fundamental issue in semi-classical quantum gravity. This paradox arises from the black hole evaporation discovered by Hawking  in the framework of quantum field theory in curved spacetime \cite{Hawking:1975vcx}. 
This discovery lends physical meaning to the concept of black hole temperature, enabling black holes to be conceptualized as genuine thermodynamic systems. It further suggests that black holes can be described in terms of quantum systems as seen from outside \cite{Almheiri:2020cfm}. Nonetheless, this revelation introduces a profound challenge: image a black hole formed from the collapse of a pure quantum state~\cite{Christodoulou:2008nj}. Upon formation, the black hole begins emitting Hawking radiation, a phenomenon wherein entangled virtual particle pairs near the black hole's event horizon—spawned from vacuum fluctuations—result in one particle escaping to infinity and its counterpart plunging into the black hole, ultimately nearing its singularity. Although these entangled particles originate in a pure state, an observer external to the black hole perceives only the outgoing particles, leading the external vacuum to appear in a mixed state. As the black hole gradually evaporates to extinction, the eventual vacuum state is also mixed. Thus, a black hole that emerges from a pure state transitions to a mixed state within the framework of semi-classical quantum gravity. This evolution conflicts with the principle of unitary evolution fundamental to quantum mechanics, culminating in the contentious issue of black hole information loss.

Addressing this challenge necessitates calculating the black hole's entanglement entropy, also known as the von Neumann entropy $S_{vN}$ or fine-grained entropy. For a quantum system, the von Neumann entropy is defined as $S_{vN}=-Tr\left\{ \rho \ln \rho \right\}$, where $\rho$ is density matrix describing the quantum state of the system. It is obvious that $S_{vN}=0$ when the state is  pure and it remains constant under unitary time evolution.  Thus, if a black hole, conceptualized as a quantum system, forms from the collapse of a pure state, it would logically follow that $S_{vN}=0$ in the final state. However, Hawking's calculations indicate that the entanglement entropy outside the black hole monotonically increases, thus contravening unitary evolution and leading to the conclusion that information within a black hole is lost.

Nevertheless, the consensus among physicists is that black holes, when viewed as quantum systems, ought to adhere to unitary evolution. This perspective was advanced by Page, who, presupposing unitary evolution for black holes, derived the temporal evolution curve of entanglement entropy—now known as the Page curve \cite{Page:1993wv,Page:2013dx}. Consequently, the information loss paradox is reframed as the challenge of reproducing the Page curve using semi-classical methods within the confines of pure gravity, thereby preserving the integrity of unitary evolution in the context of black holes.

\begin{figure}[h!]
    \begin{tabular}{cc}
\includegraphics[height=7.5cm]{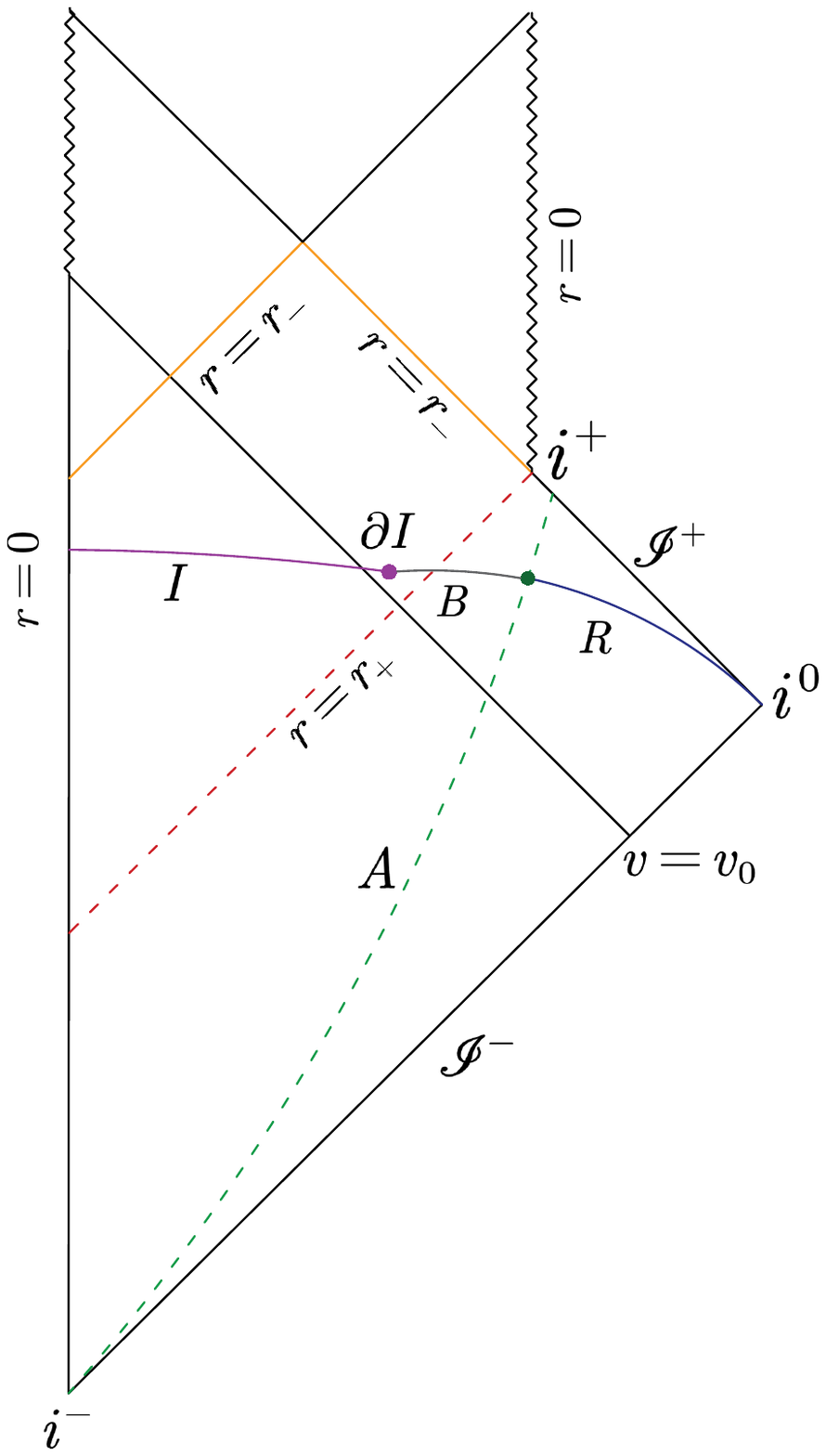}&
\includegraphics[height=7.5cm]{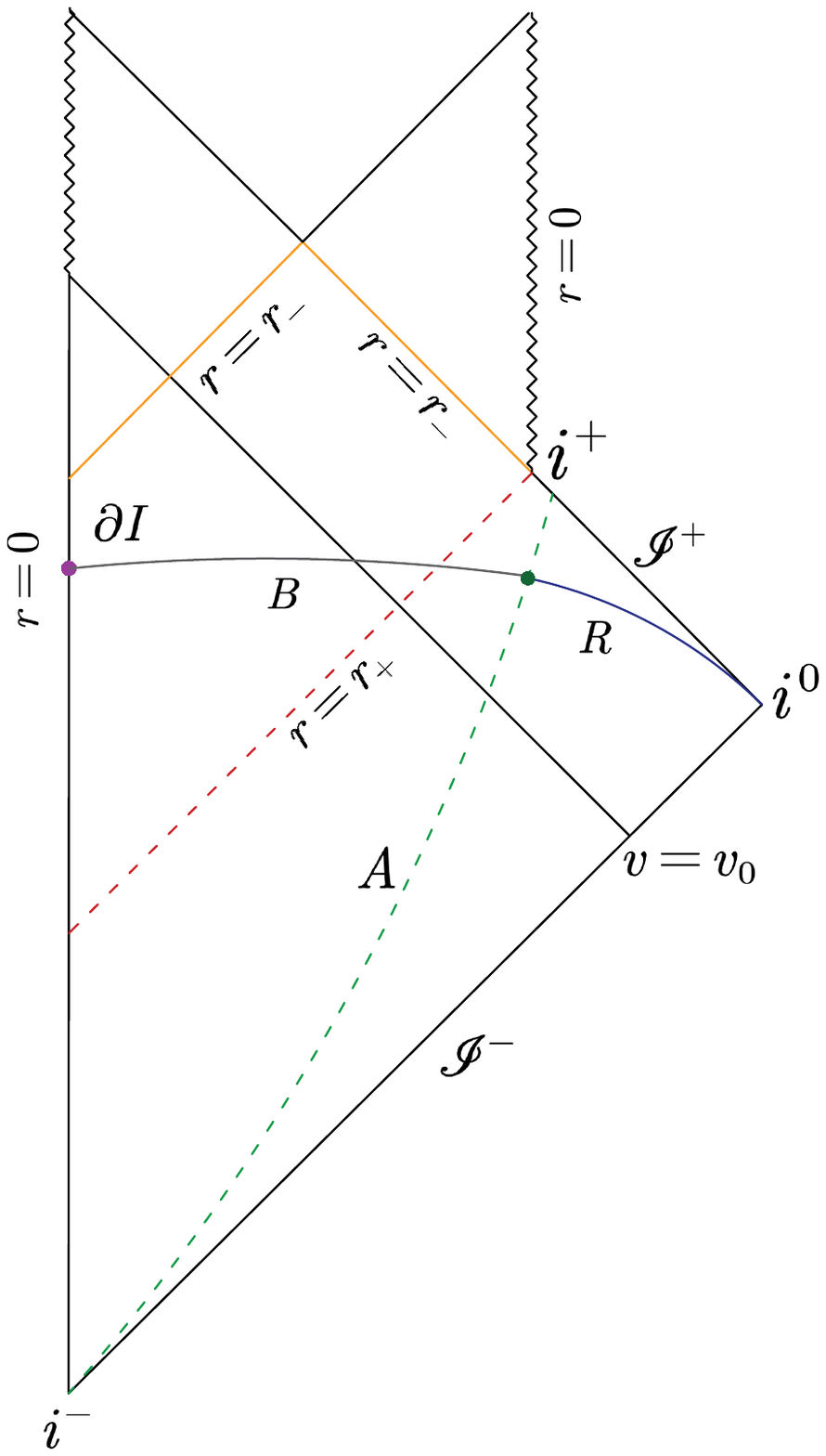}\\
	(a) & (b)  \\[6pt]
	\end{tabular}
\caption{(a) The Penrose diagram for one-sided RN black hole formed by spherical null shell collapsing at $v=v_0$. $r=r_+$(red dashed line) is called event horzion and $r=r_-$(yellow line)is usually called Cauchy horizon. Obsever on cutoff surface $A$ (green dashed line) collects Hawking radition in region $R$. Island (purple line) penetrates interior of black hole. $\partial I$ is boundary of $I$. (b) Without island, we can equivalently fix $r_I=0$.}\label{fig_Rn}
\end{figure}

The introduction of the island rule \cite{Penington:2019npb,Almheiri:2019psf,Almheiri:2019hni}, aiming to calculate the Page curve of Hawking radiation, marks a significant stride in understanding quantum effects associated with black holes. This rule posits that the fine-grained entropy of a black hole and the Hawking radiation it emits can be precisely determined by considering contributions from specific regions called islands, alongside the radiation region. The boundary of these islands, or the quantum extremal surface (QES), plays a pivotal role in extremizing the generalized entropy, encompassing both the island's area contribution and the coarse-grained entropy of matter fields.  According  to the  island rule, the fine-grained entropy of a black hole is given by:
\begin{align}\label{S_BH}
S\left( BH \right) =\min \left\{ \text{ext}\left[ \frac{\text{Area}\left( \partial I \right)}{4G_N}+S_{\text{ent}}\left( B \right) \right] \right\}, 
\end{align}
and the fine-grained entropy of Hawking radiation is 
\begin{align}\label{S_R}
S\left( R \right) =\min \left\{ \text{ext}\left[ \frac{\text{Area}\left( \partial I \right)}{4G_N}+S_{\text{ent}}\left( I\cup R \right) \right] \right\},
\end{align}
where  $R$ is the radiation region outside a cutoff surface $A$ and $I$ is called an island,  a codimension-one hypersurface extending into the internal domain of the black hole, with $\partial I$ its boundary (Fig.\ref{fig_Rn}), and $B$ is the region bounded by $\partial I$  and the cutoff surface $A$. The choice of $\partial I$ as the quantum extremal surface (QES) \cite{Engelhardt:2014gca} is based on extremizing the generalized entropy \cite{Faulkner:2013ana,Lewkowycz:2013nqa,Dong:2016hjy}
\begin{align}
S_{\text{gen}} = \frac{\text{Area}(\partial I)}{4G_N} + S_{\text{ent}},
\end{align}
where the first term quantifies the contribution from the island and the second term accounts for the coarse-grained entropy of the matter fields. To determine the fine-grained entropy of Hawking radiation in the system, the generalized entropy is calculated at all extremal points, and the minimum value is selected.
The applicability and effectiveness of the island rule have been demonstrated across various two-dimensional \cite{Almheiri:2019yqk,Gautason:2020tmk,Hartman:2020swn,Anegawa:2020ezn,Yu:2022xlh} and high-dimensional \cite{Alishahiha:2020qza,Karananas:2020fwx,Yu:2021cgi,He:2021mst,Hashimoto:2020cas,Arefeva:2021kfx,Kim:2021gzd,Wang:2021woy} gravitational models, showcasing its capacity to accurately reproduce the Page curve. 

The application of the island formula across different dimensional settings has facilitated the computation of generalized entropy, 
 $S_{\text{gen}}$, for four-dimensional asymptotically flat black holes. Historically, the majority of research on the Schwarzschild black hole \cite{Hashimoto:2020cas,Arefeva:2021kfx} and the Reissner-Nordstr\"{o}m (RN) black hole \cite{Kim:2021gzd,Wang:2021woy} has focused on eternal black holes and static vacuum configurations \cite{Matsuo:2021mmi}. In reference \cite{Alishahiha:2020qza}, one-sided dynamical black holes were discussed; however, the Hartle-Hawking (HH) state \cite{Hartle:1976tp}, more typical for eternal black holes, was employed. This methodological inconsistency was addressed in our recent work \cite{Gan:2022jay}, where we advocated for the use of the  ``in'' vacuum state \cite{Fabbri:2005mw}. This state aptly describes an asymptotically flat, one-sided dynamic black hole resulting from the collapse of a spherical null shell. We demonstrated that the $s$-wave approximation is appropriate for the ``in'' vacuum state when the cutoff surface $A$ is sufficiently distant from the horizon, thereby making it a suitable model for such dynamic black hole scenarios. This approach was later expanded to include the step-function Vaidya model of evaporating black holes in \cite{Guo:2023gfa}, further illustrating the versatility and applicability of this methodology in dynamic black hole environments.

Building upon this enhanced framework, we propose that employing the ``in'' vacuum state offers a more appropriate methodology for calculating the entanglement entropy of dynamical black holes formed from the gravitational collapse of spherical null shells. This method may extend to a broader array of black hole scenarios. Unlike Schwarzschild black holes, charged black holes exhibit dual horizons and distinct causal structures. Additionally, extremal charged black holes display unique geometric and causal configurations compared to non-extremal ones. These variations might influence the behaviors of entanglement entropy in charged black holes differently. Moreover, we intend to explore whether the strategies described in \cite{Gan:2022jay} could tackle the issue of information loss in collapsed charged black holes.

This paper conducts a thorough analysis of the island and Page curves for a charged black hole formed from the gravitational collapse of a spherical null shell, employing the island formula in the ``in'' vacuum state. We focus particularly on the effects of electrical charges on the positioning of the island and the timing of the Page curve transition. Additionally, we delve into the specific traits of extremal charged black holes. Compared to their non-extremal counterparts, extremal charged black holes present differing geometric and causal structures, necessitating tailored  analyses. We aim to demonstrate how these differences substantially influence the entropy behavior, offering new insights into the thermodynamics of these complex systems. This investigation not only broadens our understanding of black hole thermodynamics but also enhances our knowledge of the dynamics within different black hole configurations.

The paper is organized as follows. In Sec. \ref{sec_entropy},  We discuss the methodology for calculating the entanglement entropy, $S_{ent}$, in two-dimensional conformal flat spacetime, demonstrating how it varies depending on the vacuum state we select. Next, we focus on the case of non-extremal RN black holes. We analyze two scenarios: In Sec. \ref{sec_far_from}, we discuss the scenario when the cutoff surface is situated away from the horizon, we find that the boundary of the island $\partial I$ is inside and near the horizon. In Sec. \ref{sec_near}, we analyze the situation where the cutoff  surface is near the horizon, and the boundary of the island $\partial I$ appears outside and near the horizon. In both cases, the island preserves the bound of entropy. In Sec. \ref{sec_Ext}, we discuss the case of extremal RN black hole formed by the collapse of a spherical null shell and find different results compared to non-extremal RN black hole. Finally, we present our conclusion in Sec. \ref{sec_con}.

\section{Entanglement Entropy in Black Hole Background}\lb{sec_entropy}
 \renewcommand{\theequation}{2.\arabic{equation}}\setcounter{equation}{0}
 
 In this section, we establish that the $s$-wave approximation is appropriate for analyzing charged black holes, which allows for the simplification of the entanglement entropy computation to a two-dimensional spacetime framework. We also introduce a calculation method for the entanglement entropy of material fields in two-dimensional spacetime, considering different cutoff surfaces. For the scope of this paper, our focus will be on the entanglement properties associated with the ``in'' vacuum state \cite{Gan:2022jay} in the context of  a RN black hole.  We consider the union 
 $I\cup B\cup R$  as a Cauchy slice where the quantum state is pure, in alignment with our assumption of a vacuum state being pure. This setup leads to the relationship:
 
 \begin{align}\lb{equal}
S_{\text{ent}}\left( B \right) =S_{\text{ent}}\left( I\cup R \right). 
\end{align}
In the following, we will mainly focus on $S_{\text{ent}}(B)$ for simplicity.

\subsection{Cutoff surface far from horizon}

We first consider the scenario when the cutoff surface $A$ is far from the horizon.  In the context of  a RN black hole, the line element is described by
\begin{align} \lb{metric}
ds^2=-\frac{\left( r-r_+ \right) \left( r-r_- \right)}{r^2}dt^2+\frac{r^2}{\left( r-r_+ \right) \left( r-r_- \right)}dr^2+r^2\left( d\theta ^2+\sin ^2\theta d\varphi ^2 \right). 
\end{align}
where 
\begin{align}\lb{r_h_c}
r_{\pm}=GM\pm \sqrt{G^2M^2-GQ^2},
\end{align} 
$r=r_+$ is called the event horzion and $r=r_-$ is called the Cauchy horizon or inner horizon.   The effective potential in the Klein-Gordon (KG) equation for a massless scalar field is given by
\begin{align}
V_l\left( r \right) =\frac{\left( r-r_+ \right) \left( r-r_- \right)}{r^2}\left[ \frac{r_++r_-}{r^3}-\frac{2r_+r_-}{r^4}+\frac{l\left( l+1 \right)}{r^2} \right]. 
\end{align}
It is clear that $V_l\left( r \right) \rightarrow 0$ when $r \rightarrow r_+$ or $r \rightarrow \infty$.  So,  the field can be treated as an effective two-dimensional free massless field when $ r \rightarrow 2GM$ (event horizon) or $r\rightarrow \infty $, validating the approximation that higher angular momentum modes (higher $l $ values) contribute negligibly in these regions. Thus, near the event horizon and at large distances from the black hole, the field dynamics can  also be  effectively considered in two dimensions using the $s$-wave approximation as in   Schwarzschild spacetime. \cite{Gan:2022jay}

In this context, our purpose is to calculate the entanglement entropy $S_{\text{ent}}$ of a massless scalar field within a specific region in a two-dimensional ($2D$) curved spacetime.  Accurate computation of $S_{\text{ent}}$ is contingent upon the selection of an appropriate vacuum state, which in turn depends on the local coordinate system used within the region. The choice of coordinates significantly influences the methodology for calculating entanglement entropy.  Many  previous studies \cite{Alishahiha:2020qza, Hashimoto:2020cas, Arefeva:2021kfx, Kim:2021gzd, Wang:2021woy} have employed Kruskal coordinates for their analyses, which naturally led to the adoption of the Hartle-Hawking vacuum state. This state is particularly well-suited for scenarios involving eternal black holes and is often preferred for its symmetrical properties across the horizons, reflecting a thermal equilibrum environment at the Hawking temperature.
 
 However, diverging from previous approaches, this paper employs  the $(u_{in},v)$ coordinates.  These coordinates are adapted to describe the physics from the perspective of an observer falling into the black hole, which changes the nature of the vacuum state considered. Specifically, the vacuum state we select for analysis here is the ``in'' vacuum state, as defined in $(u_{in},v)$ coordinates. This vacuum state is particularly suitable for scenarios where black holes are formed dynamically, such as from collapsing matter. The ``in'' vacuum state effectively describes incoming modes from the past null infinity that are unaffected by the black hole's formation. This choice aligns with our focus on dynamic processes, contrasting with the static or equilibrium scenarios often assumed with the Hartle-Hawking state. By employing the  $(u_{in},v)$  coordinates and the corresponding ``in'' vacuum state, we aim to capture a more realistic picture of quantum field behavior in spacetimes where black holes form and evolve. This approach not only allows us to explore the properties of entanglement entropy under dynamic conditions but also enhances our understanding of the quantum fields in curved spacetime, particularly in relation to black hole information paradox and the behavior of Hawking radiation under non-equilibrium conditions.

In flat spacetime, $2D$ massless scalar fields can be effectively described by two-dimensional conformal field theory ($\text{CFT}_2$). Within this framework, the entanglement entropy $S_{\text{ent}}$  for a specific region in the vacuum state is expressed as \cite{Calabrese:2004eu,Calabrese:2009qy,Casini:2009sr}:
\begin{align}
S_{\text{ent}} = \frac{c}{3} \ln \left( d(A, I) \right),
\end{align}
where $c$ represents the central charge, and  $A $ and $I $ are the endpoints of the region under consideration. The distance $d(A,I)$  between these endpoints in flat null coordinates, where the line element is $ds^2=-dx^+dx^-$, is given by:

\begin{align}
d(A, I) = \sqrt{\left[ x^+(A) - x^+(I) \right] \left[ x^-(I) - x^-(A) \right]}.
\end{align}
 Here, the vacuum state is defined such that the vacuum expectation value (VEV) of normal ordered stress tensor in coordinates $(x^+,x^-)$ vanishes (i.e. $\left< 0 \right|:T_{x^+x^+}:\left| 0 \right> =\left< 0 \right|:T_{x^-x^-}:\left| 0 \right> =0$).

In a general $2D$ gravitational setting, it is possible to perform a Weyl transformation in local coordinates  $(x^+,x^-)$  such that the spacetime metric becomes:

\begin{align}
ds^2 = -e^{2\rho(x^+, x^-)} dx^+ dx^-,
\end{align}
where $\rho \left( x^+,x^- \right)$ is a conformal factor. The vacuum expectation value (VEV) of the normal-ordered stress tensor, defining a set of functions $t_\pm(x^\pm)$, is:

\begin{align}
\langle \Psi | :T_{\pm \pm}(x^\pm): | \Psi \rangle \equiv -\frac{\hbar}{12\pi} t_\pm(x^\pm),
\end{align}
where $|\Psi\rangle$ denotes an arbitrary quantum state. After a Weyl transformation $g \rightarrow \Omega^2 g$, the transformation of the entanglement entropy is \cite{Almheiri:2019psf}:

\begin{align}
S_{\Omega^2 g} = S_g - \frac{c}{6} \sum_{\text{endpoints}} \ln(\Omega),
\end{align}
where ``endpoints'' refers to the endpoints of the interval for which the entanglement entropy is calculated. By setting   $\Omega ^{-2}=e^{2\rho}$
  and incorporating $S_{\text{ent}}$ from above into the entanglement entropy transformation, we derive the matter part of the generalized entropy in a $2D$ spacetime \cite{Gautason:2020tmk}:

\begin{align}\lb{S_ent_far}
S_{\text{ent}} = \frac{c}{6} \ln \left( d(A, I)^2 e^{\rho_A} e^{\rho_I} \right)_{t{\pm}=0} = \frac{c}{12} \ln \left( d(A, I)^4 e^{2\rho_A} e^{2\rho_I} \right)_{t{\pm}=0},
\end{align}
where the UV cutoff parameter is omitted as it can be absorbed into the renormalization of the gravitational constant 
$G$\cite{Almheiri:2019psf,Alishahiha:2020qza,Hashimoto:2020cas,Gautason:2020tmk,Susskind:1994sm}. The subscript 
 $t_{\pm}=0$   means we should choose $\left( e^{2\rho} \right) _{t_{\pm}=0}=e^{2\rho \left( x^+,x^- \right)}$ (i.e. choose the coordinates that defines the vacuum state to calculate the conformal factor)\footnote{For example, for Hartle-Hawking state $|H\rangle$, $t_{\pm}=0$ means $\langle H|:T_{UU}:|H\rangle=\langle H|:T_{VV}:|H\rangle$=0, we need to choose Kruskal coordinates $x^+=U, x^-=V$, i.e. $\left( e^{2\rho} \right) _{t_{\pm}=0}=e^{2\rho \left( U,V \right)}$. In this paper, we use ``in''  vacuum state $|in\rangle$, then $t_{\pm}=0$ means $\langle in|:T_{u_{in}u_{in}}:|in\rangle=\langle in|:T_{vv}:|in\rangle$=0, we need to choose coordinates $x^+=u_{in}, x^-=v$, i.e. $\left( e^{2\rho} \right) _{t_{\pm}=0}=e^{2\rho \left( u_{in},v \right)}$. }.

\subsection{Cutoff surface near horizon}

When the cutoff surface is close to the horizon, the methodology for calculating the entanglement entropy of matter fields, denoted as $S_{\text{ent}}$, requires re-evaluation due to the unique spatial considerations of the region. In scenarios where the spatial region under investigation is small relative to the curvature of the surrounding spacetime, the curvature's influence on the matter fields can be considered negligible. Consequently, the area can be approximated as flat, and the entanglement entropy of the matter fields can be modeled as if in a vacuum state within flat spacetime.
This approach is particularly useful when handling physical systems near black hole horizons, where spacetime curvature is significant yet local flatness assumptions can still apply over small distances. The renormalization technique comes into play to adjust for the effects of curvature on the quantum fields, allowing us to determine the renormalized entanglement entropy effectively. This method is defined by the formula \cite{Hashimoto:2020cas,Casini:2005zv}:
\begin{align}\lb{S_ent_near}
S_{\text{ent}}=-kc\frac{A}{L^2},
\end{align}
where $k$ is a constant related to the specifics of the renormalization, $c$ represents the central charge associated with the field theory, $A$ denotes the area of the cutoff surface, and $L$ is the geodesic distance between the cutoff surface $A$ and the boundary of the island $\partial I$.

In this paper, we assume that both the island $\partial I$ and the cutoff surface $A$ are located near the horizon, then $L$ can be approximately determined by \cite{Hashimoto:2020cas}
\begin{align}
	\begin{split}
	L&=\int{\sqrt{-e^{2\rho \left( x^+,x^- \right)}dx^+dx^-}}\\
	&\approx \sqrt{\left[ x^+\left( A \right) -x^+\left( I \right) \right] \left[ x^-\left( I \right) -x^-\left( A \right) \right] e^{\rho _A}e^{\rho _I}}\\
	&=\sqrt{d\left( A,I \right) ^2 e^{\rho _A}e^{\rho _I}}.\\
	\end{split}
\end{align}
This approximation,  which is  valid when $L $ is sufficiently small with respect to the length
scale of the curvature, enables us to calculate entanglement entropy conveniently.

\section{ Cutoff surface far from horizon}\lb{sec_far_from}
 \renewcommand{\theequation}{3.\arabic{equation}}\setcounter{equation}{0}
 As mentioned earlier, the calculation of entanglement entropy of mater term depends on the position of the cutoff surface. In this section, we first focus on the entanglement entropy of non-extremal charged black holes when the cutoff surface is far from the horizon.

	\subsection{With island}
	
	In the scenario of a one-sided non-extremal RN black hole formed by the collapse of a spherical null shell at $v=v_0$, the transition between the pre-collapse and post-collapse spacetime geometries is significant for understanding the dynamics around the black hole.  The spacetime configuration represented by a Penrose diagram, referenced as Fig.\ref{fig_Rn}, provides a visual representation of the causal relationships and the general structure of the spacetime surrounding the black hole.
	
Before the collapse ($v<v_0$), the spacetime can be described  in double null coordinates, which is represented by:

\begin{align}
ds^2=-du_{in}dv+r_{in}^{2}\left( d\theta ^2+\sin ^2\theta d\varphi ^2 \right). 
\end{align}
where $u_{in}$ and $v$ are the ingoing and outgoing null coordinates respectively, and 
$r_{in}$ is the radial coordinate in this pre-collapse spacetime. This describes a flat spacetime unaffected by the gravitational fields of the black hole, as no mass-energy has yet influenced this region. Following the collapse ($v>v_0$), the spacetime geometry transitions to that of a RN black hole. In double null coordinates, the RN spacetime is given by:

\begin{align}
ds^2=-\frac{\left( r-r_+ \right) \left( r-r_- \right)}{r^2}dudv+r^2\left( d\theta ^2+\sin ^2\theta d\varphi ^2 \right), 
\end{align}
where 
\begin{align}
u_{in}=t_{in}-r_{in}\ ,\ u=t-r^*\ ,\ v=t_{in}+r_{in}=t+r^*,
\end{align}
and $ r^*$ is the tortoise coordinate defined as:

\begin{align}
r^*=r+\frac{r_{+}^{2}}{r_+-r_-}\ln \left|\frac{ r-r_+}{r_++r_-}\right|-\frac{r_{-}^{2}}{r_+-r_-}\ln \left|\frac{ r-r_-}{r_++r_-} \right|.
\end{align}
This coordinate is crucial for describing the effective ``distance'' that takes into account the gravitational warping effects near the black hole horizons.

The connecting condition ensures the metric continuity across the transition from the ``in'' region (before the shell collapses) to the ``out'' region (post-collapse), characterized by the expression:
\begin{align}
u=u_{in}+\frac{2r_{-}^{2}}{r_+-r_-}\ln \left| \frac{v_0-u_{in}-2r_-}{2(r_++r_-)} \right|-\frac{2r_{+}^{2}}{r_+-r_-}\ln \left| \frac{v_0-u_{in}-2r_+}{2(r_++r_-)} \right|.
\end{align}
At the late times, when $u \rightarrow +\infty$ , $u_{in}$ simplifies to:
\begin{align}
u_{in}\approx -2(r_++r_-)e^{-\frac{u\left( r_+-r_- \right)}{2r_{+}^{2}}}+v_0-2r_+.
\end{align}
In the Kruskal coordinates,
\begin{align}
	\begin{split}
U=-e^{-\kappa u},\ &V=e^{\kappa v}\left( \text{Outside\,\,horizon} \right) \\
U=e^{-\kappa u},\ &V=e^{\kappa v}\left( \text{Inside\,\,horizon} \right) 
	\end{split}
\end{align}
where $\kappa=\frac{r_+-r_-}{2r_+^2}$ is the surface gravity and the line element in the Kruskal coordinates becomes
\begin{align}
ds^2=\frac{f\left( r \right)}{\kappa ^2UV}dUdV+r^2\left( d\theta ^2+\sin ^2\theta d\varphi ^2 \right),
\end{align}
where $f(r)=\frac{\left( r-r_+ \right) \left( r-r_- \right)}{r^2}$. We  can  also write
\begin{align}
u_{in}\approx 2(r_++r_-)U+v_0-2r_+,
\end{align}
then
\begin{align}
\frac{dU}{du_{in}}\approx \frac{1}{2(r_++r_-)}\ ,\ \frac{dV}{dv}=\kappa V.
\end{align}
Now $ds^2=-e^{2\rho \left( u_{in},v \right)}du_{in}dv$ covers the whole spacetimes. In  the ``in'' region $v<v_0$, $e^{2 \rho (u_{in},v)}=1$, and in the ``out'' region
\begin{align}
e^{2\rho \left( u_{in},v \right)}=e^{2\rho \left( U,V \right)}\frac{dU}{du_{in}}\frac{dV}{dv} =-\frac{f\left( r \right)}{2(r_++r_-)\kappa U}.
\end{align}
The generalized entropy of the system is given by
\begin{align}
S_{\text{gen}}=S_{\text{gravity}}+S_{\text{ent}},
\end{align}
where
\begin{align}
S_{\text{gravity}}=\frac{\pi r_{I}^{2}}{G}
\end{align}
is computed from the area of the boundary of the island $\partial I$.

The $s$-wave approximation is valid since the island is far from the horizon. The ``in'' vacuum state $|in\rangle$ of  the Minkowski region $v<v_0$, defined with respect to coordinates $(u_{in},v)$, ensures that the vacuum expectation value (VEV) of the normal ordered stress tensor in these coordinates vanishes \cite{Gan:2022jay,Fabbri:2005mw}. Consequently, $t_\pm=0$ for the vacuum state $|in\rangle$, which leads to:
\begin{align}
\left( e^{2\rho} \right) _{t_{\pm}=0}=e^{2\rho \left( u_{in},v \right)}.
\end{align}
This is the critical difference between our paper and the others \cite{Alishahiha:2020qza,Hashimoto:2020cas,Arefeva:2021kfx,Kim:2021gzd,Wang:2021woy}. 

The location of the island, whether outside or inside the horizon, plays a significant role in determining the physical implications and the mathematical description of entanglement entropy. If the island is outside the horizon 
\begin{align}
	\begin{split}
U_IV_I&=-e^{\frac{r_I\left( r_+-r_- \right)}{r_{+}^{2}}}\left( \frac{r_I-r_+}{r_++r_-} \right) \left(\frac{ r_I-r_-}{r_++r_-} \right) ^{-\frac{r_{-}^{2}}{r_{+}^{2}}}\\
&\approx -e^{\frac{r_+-r_-}{r_+}} \left(\frac{r_I-r_+}{r_++r_-}\right)\left(\frac{r_+-r_-}{r_++r_-}\right)^{-\frac{r_-^2}{r_+^2} },
	\end{split}
\end{align}
where we have assumed  $r_I \approx r_+$. Then we obtain
\begin{align}
r_I\approx r_+-(r_++r_-)e^{-\frac{r_+-r_-}{r_+}}(\frac{r_+-r_-}{r_++r_-})^{-\frac{r_-^2}{r_+^2} }U_IV_I=r_+-U_IV_I\lambda ,
\end{align}
where $\lambda =(r_++r_-)e^{-\frac{r_+-r_-}{r_+}}(\frac{r_+-r_-}{r_++r_-})^{\frac{r_-^2}{r_+^2} }$.
If the island is inside the horizon 
\begin{align}
U_IV_I=e^{\frac{r_I\left( r_+-r_- \right)}{r_{+}^{2}}}\left( \frac{r_+ -r_I}{r_++r_-} \right) \left( \frac{r_I-r_-}{r_++r_-} \right) ^{-\frac{r_{-}^{2}}{r_{+}^{2}}},
\end{align}
the same formula holds:
\begin{align}
r_I\approx r_+ -U_IV_I\lambda.
\end{align}
Using the $s$-wave approximation and formula (\ref{S_ent_far}), the entanglement entropy $S_{\text{ent}}$ 
can be expressed as:
\begin{align}\lb{S_ent_}
	\begin{split}
S_{\text{ent}}&=\frac{c}{12}\ln \left\{ \left[ u_{in}\left( A \right) -u_{in}\left( I \right) \right] ^2\left[ v_I-v_A \right] ^2\frac{f\left( r_A \right) f\left( r_I \right)}{\kappa ^2U_IU_A} \right\} \\
&\approx \frac{c}{12}\ln \left[ \left( U_A-U_I \right) ^2\left( \ln \frac{V_I}{V_A} \right) ^2\frac{V_I\lambda\left( r_+-r_--U_IV_I\lambda \right)}{\kappa ^4\left( r_+-U_IV_I\lambda \right) ^2U_A} \right],
	\end{split}
\end{align}
where we have used $f(r_A) \approx 1$ since $r_A\gg r_+$.
We assume that  $\partial I $ is near the event horizon, thus $U_I \rightarrow 0$. Expanding above formula (\ref{S_ent_}) to first order of $U_I$, we obtain 
\begin{align}
S_{\text{ent}}=\frac{c}{12}\left\{ \ln \left| \frac{V_IU_A\lambda\left( r_+-r_- \right)}{r_{+}^{2}\kappa ^4}\left( \ln \frac{V_I}{V_A} \right) ^2 \right|+\frac{r_+-2r_-}{r_+\left( r_+-r_- \right)}U_IV_I\lambda-2\frac{U_I}{U_A}+\mathcal{O}\left( U_{I}^{2} \right) \right\}.
\end{align}
 Similarly, for the gravitational entropy, we derive:
\begin{align}
S_{\text{gravity}}=\frac{\pi}{G}\left( r_+-U_IV_I\lambda\right) ^2=\frac{\pi r_{+}^{2}}{G}-\frac{2\pi U_IV_I\lambda r_+}{G}+\mathcal{O}\left( U_{I}^{2} \right). 
\end{align}
This leads to the generalized entropy:
\begin{align}\lb{S_gen_far}
S_{\text{gen}}=\frac{\pi r_{+}^{2}}{G}+\frac{c}{12}\ln \left| \frac{V_IU_A\lambda\left( r_+-r_- \right)}{r_{+}^{2}\kappa ^4}\left( \ln \frac{V_I}{V_A} \right) ^2 \right|+\frac{c}{12}U_I\left[ V_I\lambda\left( \frac{r_+-2r_-}{r_+\left( r_+-r_- \right)}-\frac{24\pi r_+}{cG} \right) -\frac{2}{U_A} \right] +\mathcal{O}\left( U_{I}^{2} \right). 
\end{align}
Extremizing above formula (\ref{S_gen_far}) with respect to $U_I$ , we find:
\begin{align}
\frac{\partial S_{\text{gen}}}{\partial U_I}\approx V_I\lambda\left( \frac{r_+-2r_-}{r_+\left( r_+-r_- \right)}-\frac{24\pi r_+}{cG} \right) -\frac{2}{U_A}=0,
\end{align}
which yields the solution:
\begin{align}
V_I=\frac{2cG\left( r_+-r_- \right) r_+}{\lambda U_A\left[ \left( r_+-2r_- \right) cG-24\pi r_{+}^{2}\left( r_+-r_- \right) \right]}.
\end{align}
Further extremizing with respect to $V_I$, we get
\begin{align}\lb{PS_PV}
\frac{\partial S_{\text{gen}}}{\partial V_I}\approx \frac{1+\frac{2}{\ln \frac{V_I}{V_A}}}{V_I}+U_I\lambda \left( \frac{r_+-2r_-}{r_+\left( r_+-r_- \right)}-\frac{24\pi r_+}{cG} \right) =0.
\end{align}
Dropping $\ln\frac{V_I}{V_A}$  in eq.(\ref{PS_PV}) since $\ln\frac{V_I}{V_A}\ll 1$, we can derive
\begin{align}
U_I=-\frac{U_A}{2}.
\end{align}
Hence, $U_I<0$, which means $\partial I$  is inside the horizon. In contrast, the location of the island derived using the HH vacuum state is outside the event horizon, as discussed \cite{Wang:2021woy,Tong:2023nvi}. This is significantly different from our calculated results. In addition,
\begin{align}\lb{UV}
U_IV_I=-\frac{cG\left( r_+-r_- \right) r_+}{\lambda\left[ \left( r_+-2r_- \right) cG-24\pi r_{+}^{2}\left( r_+-r_- \right) \right]}\approx \frac{cG}{24\pi \lambda r_{+}}\ll 1,
\end{align}
where we have used $cG/\lambda r_{+}\ll 1$. As a consequence, $r_{I}^{*} \rightarrow -\infty$ as $U_IV_I \rightarrow 0$, thus confirming that $r_I \approx r_+$. When $r_- = 0$ (i.e. $Q = 0$), the RN black hole reduces to a Schwarzschild black hole, leading to similar conclusions as in the article \cite{Gan:2022jay}.

\begin{figure}[h!]
	\centering
	\includegraphics[height=7.5cm]{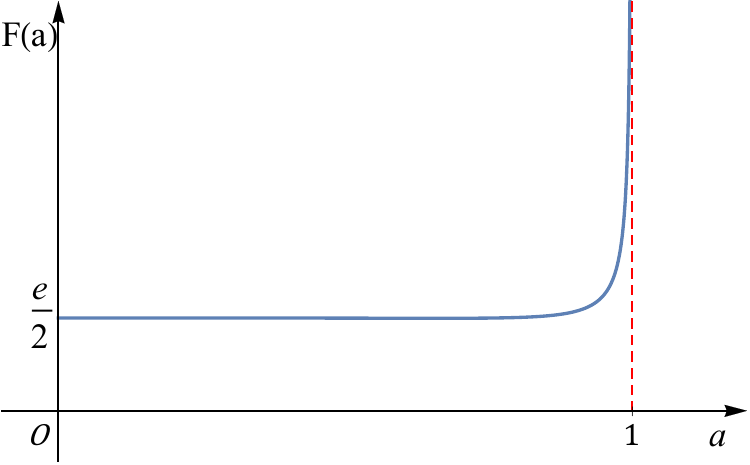}
	\caption{The behavior of $F(a)$, where $a=Q/\sqrt{G}M$. When $a$ is not close to $1$, the value of $F(a)$ remains relatively constant.}
\label{Fa}
\end{figure}

By substituting eq.(\ref{r_h_c}) into eq.(\ref{UV}) and setting $Q = a\sqrt{G} M$, where $a$ is a dimensionless constant, eq.(\ref{UV}) becomes:
\begin{align}
U_IV_I=\frac{c}{48\pi GM^2}F(a),
\end{align}
where
\begin{align}
F(a)=\frac{(1-a^2)^{-\frac{(1-\sqrt{1-a^2})^2}{2(1+\sqrt{1-a^2})^2}}}{1+\sqrt{1-a^2}} \exp[\frac{2\sqrt{1-a^2}}{1+\sqrt{1-a^2}}].
\end{align}
 For $a = 0$, $F(0) = e/2$, corresponding to Schwarzschild black holes. As $a$ approaches 1 (near-extremal RN black hole), $F(a)$ increases indefinitely. This suggests that the method for determining the island's position may become inapplicable as $U_IV_I$ is no longer negligible. In a more general sense, the plot of $F(a)$ is shown in Fig.\ref{Fa}.

Plugging these solutions back into eq.(\ref{S_gen_far}), we find:
\begin{align}
	\begin{split}
S_{\text{gen}}&=\frac{\pi r_{+}^{2}}{G}+\frac{c}{12}\ln \left| \frac{2cG\left( r_+-r_- \right)}{r_+\kappa ^4\left[ \left( r_+-2r_- \right) cG-24\pi r_{+}^{2}\left( r_+-r_- \right) \right]}\left( \ln \frac{2cG\left( r_+-r_- \right) r_+}{\lambda V_AU_A\left[ \left( r_+-2r_- \right) cG-24\pi r_{+}^{2}\left( r_+-r_- \right) \right]} \right) ^2 \right|+\mathcal{O}\left( G \right)\\
&\approx \frac{\pi r_{+}^{2}}{G}=S_{\text{BH}},
	\end{split}
\end{align}
where the last line also uses $cG/\lambda r_+ \ll 1$. Thus, with island configurations, the late-time entropy of Hawking radiation is bounded by the black hole's Bekenstein-Hawking entropy, which decreases monotonically as the black hole evaporates.

\subsection{Without island}

In the scenario where  there is no island (i.e. $r_I=0$), the gravitational entropy $S_{\text{gravity}}$ vanishes, and $e^{2\rho \left( u_{in},v \right) \left( I \right)}=1$. Thus, the generalized entropy can be expressed as follows:
\begin{align}\lb{S_gen_tA}
	\begin{split}
S_{\text{gen}}&=\frac{c}{12}\ln \left| \left( 2U_A+v_0-2r_+-t_I \right) ^2\left( v_I-v_A \right) ^2\frac{f\left( r_A \right)}{2\kappa U_A} \right|\\
&=\frac{c}{6}\ln \left| -2e^{-\kappa \left( t_A-r_{A}^{*} \right)}+v_0-2r_+-t_I \right|+\frac{c}{6}\ln \left| t_I-t_A-r_{A}^{*} \right|+\frac{c}{12}\ln \left| \frac{f\left( r_A \right)}{2\kappa} \right|-\frac{c}{12}\ln \left| -e^{-\kappa \left( t_A-r_{A}^{*} \right)} \right|\\
&\approx \frac{c}{12}\kappa t_A,
	\end{split}
\end{align}
where in the last line we have used the limit $t_A\gg r_A\gg r_+$.  This implies that, without an island, the radiation entropy at late times grows linearly with time, which aligns with Hawking's original result.
According to the island formula as given in eq.\eqref{S_R},
\begin{align}
S\left( R \right) =\min \left( S_{\text{gen}} \right) \approx \min \left( S_{\text{ent}},S_{\text{BH}} \right).
\end{align}
Thus, the Page time, which indicates the transition point at which the entropy of the radiation begins to decrease as the black hole evaporates, can be approximated by:
\begin{align}\lb{page_time}
t_{page}\approx \frac{12}{c\kappa}S_{\text{BH}}.
\end{align}
where the influence of the charge $Q$  on the Page time is reflected through its effect on the surface gravity 
$\kappa$ and the black hole entropy $S_{BH}$. The Page time we calculated in the ``in'' vacuum state is twice that of the result obtained using the HH vacuum state in \cite{Wang:2021woy,Tong:2023nvi}. As pointed out by the authors of \cite{Gan:2022jay}, the entanglement entropy $S_{\text{ent}}$ in the  ``in'' vacuum state is smaller than that  in the thermal HH state. Consequently, the Page time in the "in" vacuum state is reached later. This indicates that the calculation results vary when different vacuum states are considered.

\section{Cutoff Surface Near the Horizon}\lb{sec_near}
	\renewcommand{\theequation}{4.\arabic{equation}}\setcounter{equation}{0}
 
In this section, we consider the situation where the cutoff surface is near the horizon. Our analysis adopts an analytical approach akin to that presented in \cite{Hashimoto:2020cas}, which focuses on the Hartle-Hawking state within an eternal Schwarzschild black hole. Contrarily, our study is based on the ``in'' vacuum state within a one-sided charged black hole formed by the collapse of a null shell. Despite these contextual differences, when the region of interest is considerably smaller than the scale of curvature, the entropy of the matter fields can effectively be approximated by that in flat spacetime. This approximation holds true irrespective of the spacetime  or the state under consideration.

\subsection{With island}

When considering scenarios with an island, we still have 
\begin{align}
S_{\text{gravity}}=\frac{\pi r_{+}^{2}}{G}.
\end{align}
Given that both the island boundary, $\partial I$, and the cutoff surface $A$ are assumed to be near the horizon, the effective distance, $L$, can be approximated as:
\begin{align}
	\begin{split}
L&\approx \sqrt{d\left( A,I \right) ^2e^{\rho _A}e^{\rho _I}}\\
&=\frac{1}{\kappa}\sqrt{\left( U_A-U_I \right) \left( V_I-V_A \right) \sqrt{f\left( r_A \right) f\left( r_I \right)}e^{-\kappa \left( r_{A}^{*}+r_{I}^{*} \right)}},
	\end{split}
\end{align}
then using eq.(\ref{S_ent_near}) we have
\begin{align}\lb{S_gen_near}
	\begin{split}
S_{\text{gen}}&=\frac{\pi r_{I}^{2}}{G}-\frac{4\pi kc\kappa ^2r_{A}^{2}e^{\kappa \left( r_{A}^{*}+r_{I}^{*} \right)}}{\left( U_A-U_I \right) \left( V_I-V_A \right) \sqrt{f\left( r_A \right) f\left( r_I \right)}}\\
&=\frac{\pi r_{I}^{2}}{G}-\frac{4\pi kc\kappa ^2r_Ir_{A}^{3}e^{\kappa \left( r_{A}^{*}+r_{I}^{*} \right)}}{\left( -e^{-\kappa \left( t_A-r_{A}^{*} \right)}+e^{-\kappa \left( t_I-r_{I}^{*} \right)} \right) \left( e^{\kappa \left( t_I+r_{I}^{*} \right)}-e^{\kappa \left( t_A+r_{A}^{*} \right)} \right) \sqrt{\left( r_A-r_+ \right) \left( r_A-r_- \right) \left( r_I-r_+ \right) \left( r_I-r_- \right)}}.
	\end{split}
\end{align}
Extremizing above formula (\ref{S_gen_near}) over $t_I$, we obtain 
\begin{align}
\frac{\partial S_{\text{gen}}}{\partial t_I}=-\frac{4\pi kc\kappa ^3r_Ir_{A}^{3}}{\sqrt{\left( r_A-r_+ \right) \left( r_A-r_- \right) \left( r_I-r_+ \right) \left( r_I-r_- \right)}}\frac{e^{\kappa \left( t_I+t_A \right)}\left( e^{2\kappa t_I}-e^{2\kappa t_A} \right)}{\left[ e^{\kappa \left( t_I-r_{I}^{*} \right)}-e^{\kappa \left( t_A-r_{A}^{*} \right)} \right] ^2\left[ e^{\kappa \left( t_I+r_{I}^{*} \right)}-e^{\kappa \left( t_A+r_{A}^{*} \right)} \right] ^2}=0,
\end{align}
which has solutions 
\begin{align}
t_I=t_A+\frac{c_1\pi i}{\kappa}\ ,\ c_1\in \mathbb{Z}.
\end{align}
where $\mathbb{Z}$ represents integers. We only take the real number solution
\begin{align}\lb{t_I}
t_I=t_A.
\end{align}
Putting eq.(\ref{t_I}) into eq.(\ref{S_gen_near}), then we arrive at
\begin{align}\lb{S_gen}
S_{\text{gen}}=\frac{\pi r_{I}^{2}}{G}-\frac{4\pi kc\kappa ^2r_Ir_{A}^{3}e^{\kappa \left( r_{A}^{*}+r_{I}^{*} \right)}}{\left( e^{\kappa r_{I}^{*}}-e^{\kappa r_{A}^{*}} \right) ^2\sqrt{\left( r_A-r_+ \right) \left( r_A-r_- \right) \left( r_I-r_+ \right) \left( r_I-r_- \right)}}.
\end{align}
Since we have already assumed that both $\partial I$ and the cutoff surface $A$ are near the horizon, we can use 
\begin{align}
r_A=r_+\left( 1+\alpha \right), r_I=r_+\left( 1+\beta \right), 
\end{align}
where $\beta <\alpha \ll 1$. Then expanding eq.(\ref{S_gen}) to first order of $\beta$, we have 
\begin{align}\lb{S_gen_ab}
	\begin{split}
S_{\text{gen\,\,}}=&\frac{\pi r_{+}^{2}}{G}-\frac{4\kappa ^2\pi ckr_{+}^{3}e^{-\kappa r_+\alpha}\left( 1+\alpha \right) ^3}{\alpha \left( r_+-r_- \right) \sqrt{1+\frac{\alpha}{2r_+\kappa}}}-\frac{2\sqrt{\beta}\left( 4\kappa ^2\pi ckr_{+}^{3}e^{-2\kappa r_+\alpha}\left( 1+\alpha \right) ^3 \right)}{\alpha ^{\frac{3}{2}}\left( r_+-r_- \right) \sqrt{1+\frac{\alpha}{2r_+\kappa}}}\\
&+\beta \left\{ \frac{2\pi r_{+}^{2}}{G}-\frac{4\kappa ^2\pi ckr_{+}^{3}e^{-3\kappa r_+\alpha}\left[ e^{2\kappa r_+\alpha}\left( \alpha -\frac{1}{4\kappa r_+}+\kappa r_+ \right) +3 \right]}{\alpha ^2\left( r_+-r_- \right) \sqrt{1+\frac{\alpha}{2r_+\kappa}}} \right\} +O\left( \beta ^{\frac{3}{2}} \right).
	\end{split}
\end{align}
This  result reduces to that in \cite{Gan:2022jay} when $r_-=0$.
Extremizing eq.(\ref{S_gen_ab}) over $\beta$ leads to 
\begin{align}
\frac{\partial S_{\text{gen}}}{\partial \beta}=\frac{2\pi r_{+}^{2}}{G}-\frac{4\kappa ^2\pi ckr_{+}^{3}e^{-3\kappa r_+\alpha}\left[ e^{2\kappa r_+\alpha}\left( \alpha -\frac{1}{4\kappa r_+}+\kappa r_{r_1} \right) +3 \right]}{\alpha ^2\left( r_+-r_- \right) \sqrt{1+\frac{\alpha}{2r_+\kappa}}}-\frac{4\kappa ^2\pi ckr_{+}^{3}e^{-2\kappa r_+\alpha}\left( 1+\alpha \right) ^3}{\sqrt{\beta}\alpha ^{\frac{3}{2}}\left( r_+-r_- \right) \sqrt{1+\frac{\alpha}{2r_+\kappa}}}=0,
\end{align}
which has the solution
\begin{align}
\beta \approx \frac{16\alpha \kappa ^4c^2G^2k^2r_{+}^{6}e^{-4\kappa r_+\alpha}\left( 1+\alpha \right) ^6}{\left[ 2r_{+}^{2}\alpha ^2\left( r_+-r_- \right) \sqrt{1+\frac{\alpha}{2r_+\kappa}} \right] ^2}\approx \frac{c^2G^2k^2\left( r_+-r_- \right)}{4\alpha ^3r_{+}^{5}}\ll 1.
\end{align}
This formula confirms that $\partial I$ remains close to and just outside the horizon, consistent with the assumption that $c^2 G^2 \ll r_{+}^4$. Consequently, the generalized entropy approximates the Bekenstein-Hawking entropy:

\begin{align}\lb{S_gen_}
    \begin{split}
S_{\text{gen}}&=\frac{\pi r_{+}^{2}}{G}-\frac{4\kappa ^2\pi ckr_{+}^{3}e^{-\kappa r_+\alpha}\left( 1+\alpha \right) ^3}{\alpha \left( r_+-r_- \right) \sqrt{1+\frac{\alpha}{2r_+\kappa}}}+\mathcal{O}\left( G \right)\\
&\approx S_{BH}.
    \end{split}
\end{align}

\subsection{Without island}

In scenarios lacking an island, we calculate $S_{\text{ent}}$ using eq.(\ref{S_ent_far}). This approach is justified by the distance between $r_I = 0$ and the horizon, combined with the proximity of the cutoff surface to the horizon for macroscopic black holes (where $r_+ \gg \ell_{p}^{2}$). Consequently, the resultant entropy of the system without an island, for a cutoff surface near the horizon, aligns with eq.(\ref{S_gen_tA}). Upon comparison with eq.(\ref{S_gen_}), we observe consistent results, specifically regarding the Page time:

\begin{align}
t_{page} \approx \frac{12}{c\kappa}S_{\text{BH}}.
\end{align}
This demonstrates that, in the absence of an island, the Page time, indicative of when the entropy of the radiation begins to decline, remains aligned with the predictions of traditional Hawking radiation models, emphasizing the universality of this temporal marker across different configurations.	

\section{Extremal RN Black Hole}\lb{sec_Ext}
	\renewcommand{\theequation}{5.\arabic{equation}}\setcounter{equation}{0}

\begin{figure}[h!]
    \begin{tabular}{cc}
\includegraphics[height=7.5cm]{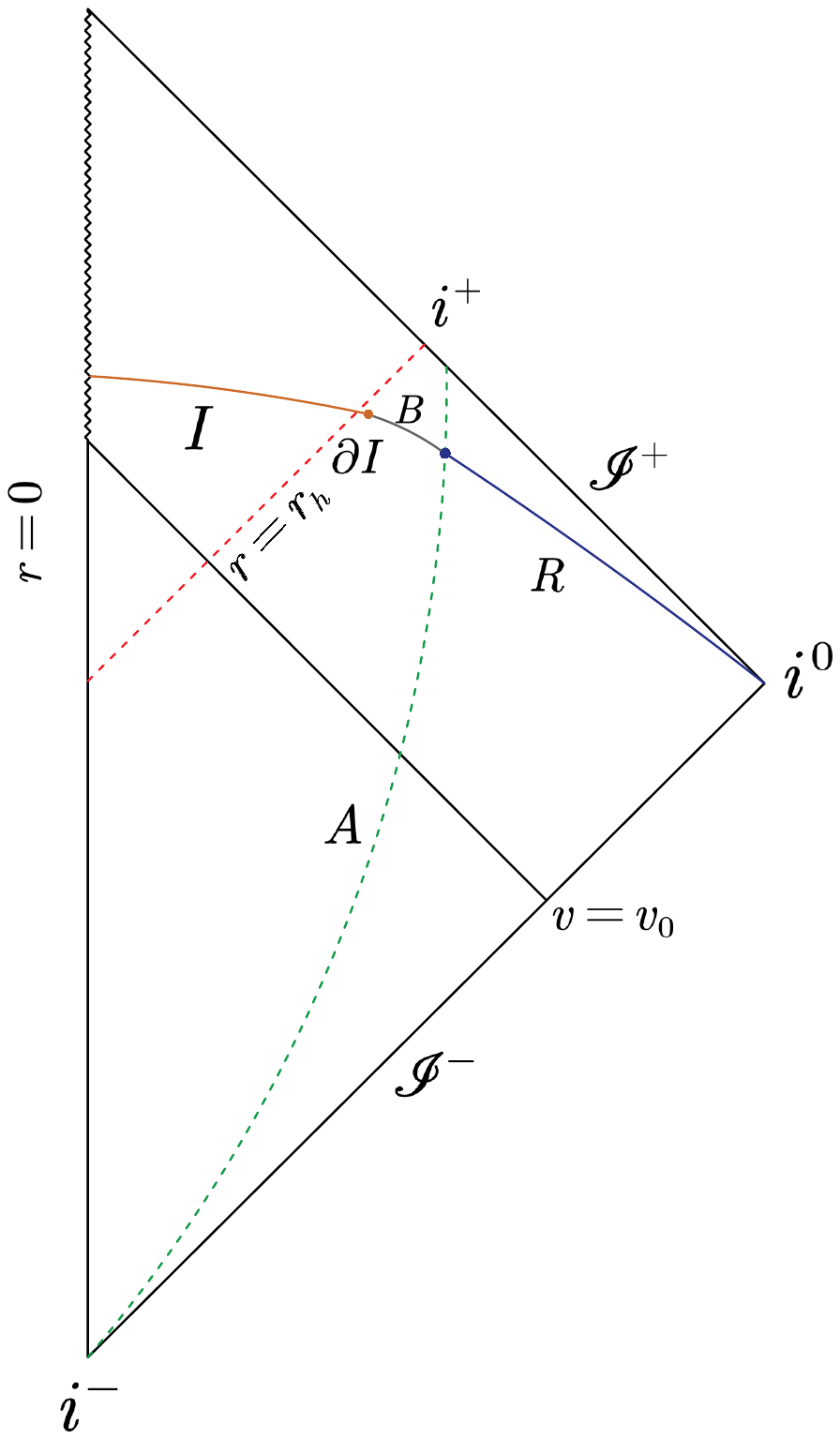}&
\includegraphics[height=7.5cm]{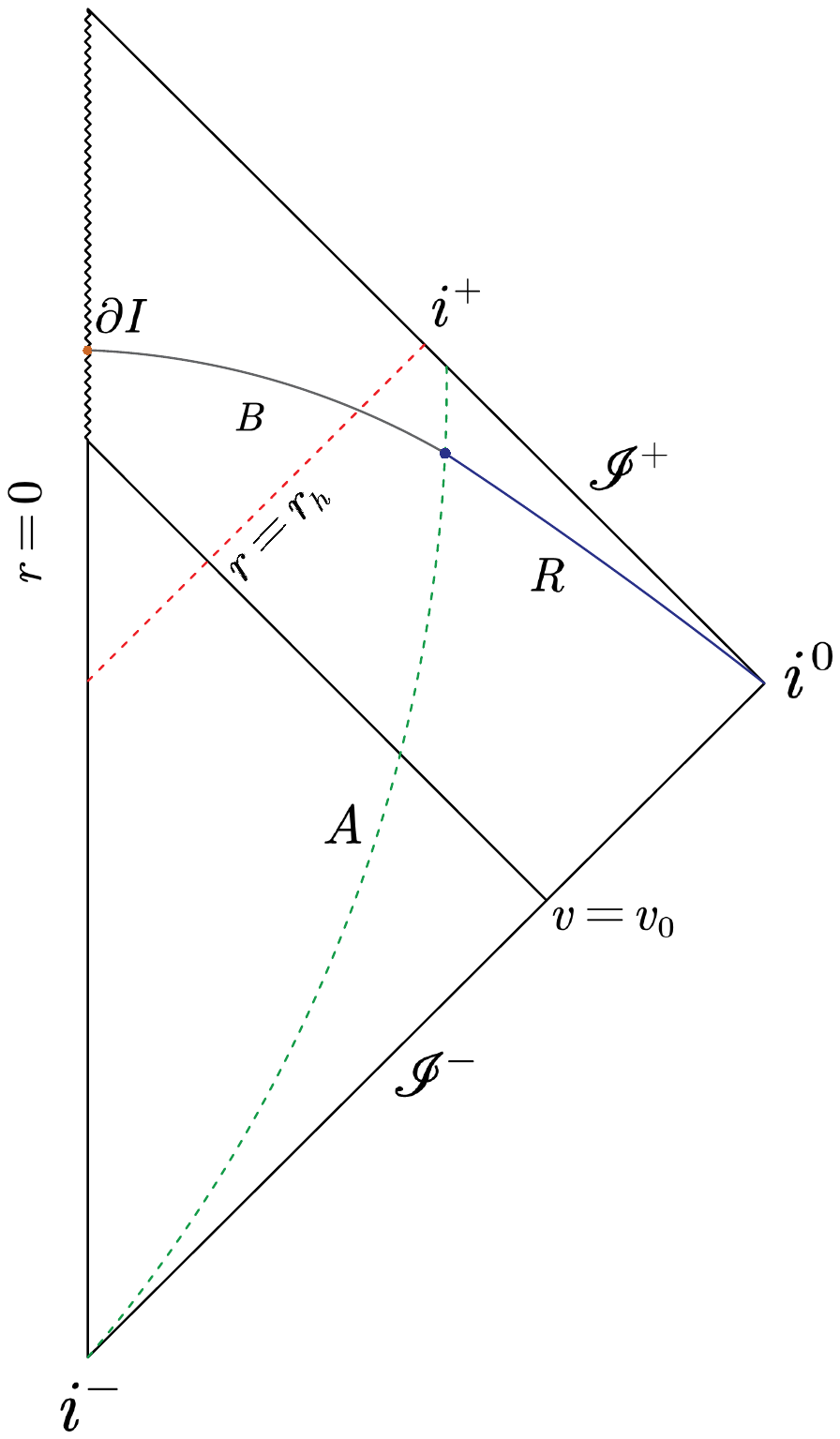}\\
	(a) & (b)  \\[6pt]
	\end{tabular}
\caption{(a) The Penrose diagram for an extremal RN black hole formed by spherical null shell collapsing at $v=v_0$. Observer on cutoff surface $A$ (green dashed line) collects Hawking radiation in region $R$.  $\partial I $ is the boundary of $I$, which is called quantum extremal surface. (b) Without island, we can equivalently fix $r_I=0$.}\label{fig_Ext}
\end{figure}

In this section, our focus is on the extremal RN black hole. Previous studies \cite{Kim:2021gzd,Ahn:2021chg} have highlighted the challenge in calculating the entanglement entropy of the extremal black hole by directly taking the extremal limit of the entanglement entropy of the non-extremal black hole. This limitation arises due to the non-continuity of the Penrose diagram between the extremal and non-extremal cases. Our previous findings, as depicted in Fig.\ref{Fa}, further support this observation. The Penrose diagram for one-sided external RN black hole formed by spherical null shell collapsing is given by Fig.\ref{fig_Ext}. We will also use ``in'' vacuum state to calculate $S_{\text{gen}}$.

The extremal condition is defined by $\sqrt{G}M=Q$, where the event horizon coincides with the Cauchy horizon, such that $r_+=r_-\equiv r_h$. Under this condition, the metric (\ref{metric}) simplifies to:
\begin{align}\lb{metric_Ext}
ds^2=-f(r)dt^2+\frac{1}{f(r)}dr^2+r^2\left( d\theta ^2+\sin ^2\theta d\varphi ^2 \right),
\end{align}
where $f(r)=\left( 1-\frac{r_h}{r} \right)^2$and the tortoise coordinate becomes
\begin{align}
r^*=\int{\frac{1}{f\left( r \right)}dr}=r+2r_h\ln \left| \frac{r-r_h}{r_h} \right|-\frac{r_{h}^{2}}{r-r_h}.
\end{align}
The double null coordinates are
\begin{align}
u=t-r^*,v=t+r^* \ \ \left( \text{Outside shell} \right),\\ 
u_{in}=t-r_{in},v=t+r_{in}\ \ \left( \text{Inside shell} \right). 
\end{align}
 The ``in'' region is also described by the Minkowski line element in the double null coordinates
\begin{align}
ds^2=-du_{in}dv+r^2\left( d\theta ^2+\sin ^2\theta d\varphi ^2 \right). 
\end{align}
In the ``out'' region, it is an extremal RN black hole with metric (\ref{metric_Ext}), which, in the double null coordinates, becomes
\begin{align}
ds^2=-f(r)dudv+r^2\left( d\theta ^2+\sin ^2\theta d\varphi ^2 \right).
\end{align}
The condition connecting these regions is:
\begin{align}\lb{condition}
u=u_{in}-4r_h\ln \left| \frac{v_0-u_{in}-2r_h}{2r_h} \right|+\frac{4r_{h}^{2}}{v_0-u_{in}-2r_h}.
\end{align}
At late times, two scenarios emerge:
\begin{enumerate}
\item For the case outside horizon (i.e. $u_{in}<v_0-2r_h$).  At late times where $u \rightarrow +\infty$, the last term in eq.(\ref{condition}) becomes dominant:
\begin{align}\lb{u_in}
u_{in}\left( u\right) \approx v_0-2r_h-\frac{4r_{h}^{2}}{u}.
\end{align}

\item For the case inside horizon (i.e. $u_{in}>v_0-2r_h$).  Assuming the island is near the horizon ($u_{in}\gtrapprox v_0-2r_h$), the third term remains dominant in eq.(\ref{condition}). The functional form of $u_{in}(u)$ mirrors that of the outside case, except that $u \rightarrow -\infty$:
\begin{align}\lb{eq5.19}
u_{in}\left( u \right) \approx v_0-2r_h-\frac{4r_{h}^{2}}{u}.
\end{align}
\end{enumerate}

Subsequently, the derivative and scaling transformation becomes:
\begin{align}
\frac{du}{du_{in}}=\frac{u^2}{4r_{h}^{2}},
\end{align}
and
\begin{align}
e^{2\rho \left( u_{in},v \right)}=e^{2\rho \left( u,v \right)}\frac{du}{du_{in}}=\left( 1-\frac{r_h}{r} \right) ^2\frac{\left( t-r^* \right) ^2}{4r_{h}^{2}}.
\end{align}

	\subsection{Cutoff Surface Far from Horizon}
Just like non-extremal charged black holes, the entanglement entropy of extremal charged black holes also depends on the position of the cutoff surface. 
For an extremal black hole with the cutoff surface positioned far from the horizon, using formula (\ref{S_ent_far}), we obtain
\begin{align}\lb{S_gen_far_Ext}
S_{\text{gen}}=\frac{\pi r_{I}^{2}}{G}+\frac{c}{6}\left\{ \ln \left| t_I-r_{I}^{*}-t_A+r_{A}^{*} \right|+\ln \left| t_I+r_{I}^{*}-t_A-r_{A}^{*} \right|+\ln \left| 1-\frac{r_h}{r_A} \right|+\ln \left| 1-\frac{r_h}{r_I} \right| \right\}. 
\end{align}
Extremizing eq.(\ref{S_gen_far_Ext}) over $t_I$, we find
\begin{align}
\frac{\partial S_{\text{gen}}}{\partial t_I}=\frac{c}{6}\left( \frac{1}{t_I-r_{I}^{*}-t_A+r_{A}^{*}}+\frac{1}{t_I+r_{I}^{*}-t_A-r_{A}^{*}} \right) =0,
\end{align}
which gives
\begin{align}\lb{t_I_Ext}
t_I=t_A.
\end{align}
Since $r_I$ is near the horizon, we can  express
\begin{align}\lb{r_I_Ext}
r_I=r_h\left( 1+\alpha \right), 
\end{align}
where $\left| \alpha \right|\ll 1$. Substituting this into the generalized entropy equation simplifies it to:
\begin{align}\lb{S_gen_Ext}
S_{\text{gen}}=\frac{\pi r_{I}^{2}}{G}+\frac{c}{6}\left\{ 2\ln \left| r_{I}^{*}-r_{A}^{*} \right|+\ln \left| 1-\frac{r_h}{r_A} \right|+\ln \left| 1-\frac{r_h}{r_I} \right| \right\}. 
\end{align}
By extremizing eq.(\ref{S_gen_Ext}) over $r_I$ and using eq.(\ref{r_I_Ext}), we arrive at
\begin{align}
\frac{\partial S_{\text{gen}}}{\partial r_I}=\frac{2\pi r_h\left( 1+\alpha \right)}{G}+\frac{c}{3\left[ r_h\left( 1+\alpha \right) +2r_h\ln \left| \alpha \right|-\frac{r_h}{\alpha} \right] -3r_{A}^{*}}\left( 1-\frac{1}{ 1+\alpha} \right) ^{-2}+\frac{c}{6r_h\alpha \left( 1+\alpha \right)}=0.
\end{align}
Simplifying the above equation and  leaving only the first-order term of $\alpha$ lead to
\begin{align}
12\pi \alpha r_{h}^{3}-2cGr_h\left( 1+3\alpha \right) +cG\left( r_h+\alpha r_{A}^{*} \right) =0.
\end{align}
By solving for $\alpha$, we obtain
\begin{align}
\alpha \approx \frac{cGr_h}{12\pi r_{h}^{3}-6cGr_h+cGr_{A}^{*}}\approx \frac{cGr_h}{12\pi r_{h}^{3}+cGr_{A}^{*}}\ll 1,
\end{align}
where we have used $cG\ll r_{h}^{2}$ and $r_A \gg r_h$. Obviously, $\alpha >0$,  confirming that the island is slightly outside the horizon. The final entropy expression simplifies to match the Bekenstein-Hawking entropy:
\begin{align}
	\begin{split}
S_{\text{gen}}&\approx \frac{\pi r_{h}^{2}}{G}+\frac{c}{6}\left\{ 2\ln \left| r_h\left( 1-\frac{cG}{12\pi r_{h}^{2}} \right) +2r_h\ln \left| \frac{cG}{12\pi r_{h}^{2}} \right|+\frac{12\pi r_{h}^{3}}{cG}-r_{A}^{*} \right|+\ln \left| 1-\frac{r_h}{r_A} \right|+\ln \left| 1+\frac{12\pi r_{h}^{2}}{cG} \right| \right\}\\
&\approx \frac{\pi r_h^2}{G}=S_{\text{BH}}.
	\end{split}
\end{align}

\subsection{Cutoff Surface Near Horizon}

\begin{figure}[h!]
	\centering
\includegraphics[height=7.5cm]{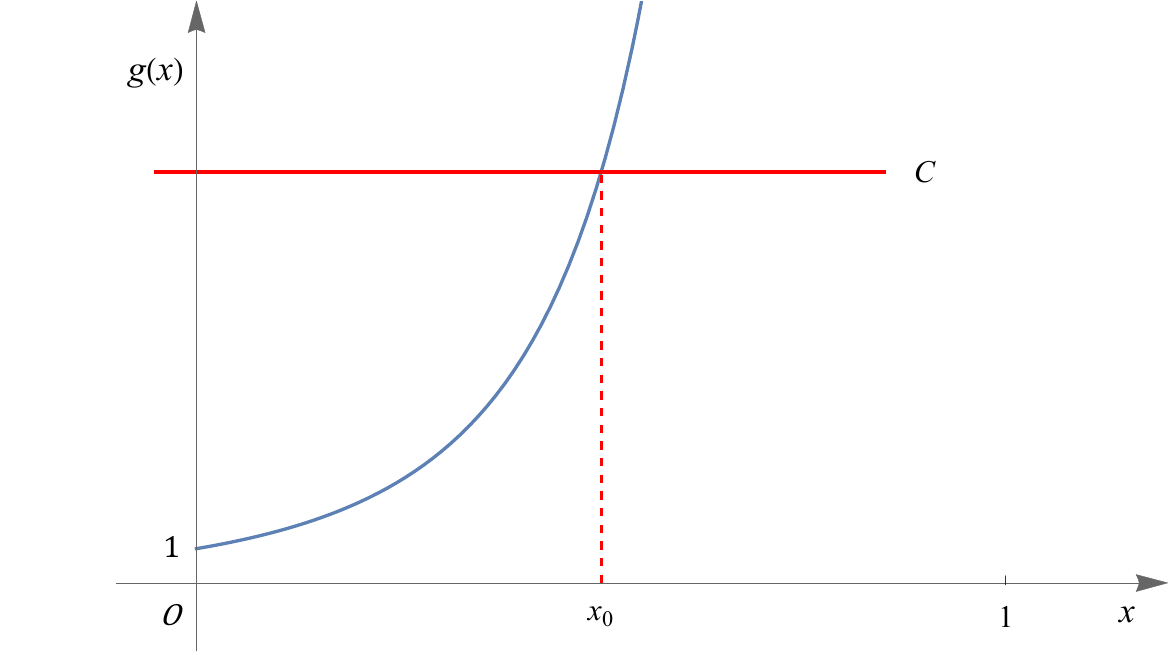}	\caption{The behavior of $g(x)$, where $g\left( x \right) =\frac{1+x}{\left( 1-x \right) ^3}$ and $C=\frac{\sigma r_{h}^{2}}{2Gkc\left( 1+\sigma \right) ^3}$.}\label{g_x}
\end{figure}

When cutoff surface is near the horizon, we use eq.(\ref{S_ent_near}) to calculate the entanglement entropy
\begin{align}
S_{\text{ent}}=-kc\frac{A}{L^2},
\end{align}
where $L$ is the geodesic distance between  the cutoff surface 
 $A$  and the island $I$. Since we assume that  both the island and the cutoff surface are near the horizon, $L$ can be approximately given by
\begin{align}
	\begin{split}
L&\approx \sqrt{d\left( A,I \right) ^2e^{\rho _A}e^{\rho _I}}\\
&=\sqrt{\left| \left[ u_{in}\left( I \right) -u_{in}\left( A \right) \right] \left( v_A-v_I \right) \left( 1-\frac{r_h}{r_I} \right) \left( 1-\frac{r_h}{r_A} \right) \frac{\left( t_I-r_{I}^{*} \right) \left( t_A-r_{A}^{*} \right)}{4r_{h}^{2}} \right|}\\
&=\sqrt{\left| \left[ -t_A+r_{A}^{*}+\left( t_I-r_{I}^{*} \right) \right] \left[ t_A+r_{A}^{*}-\left( t_I+r_{I}^{*} \right) \right] \frac{\left( r_I-r_h \right) \left( r_A-r_h \right)}{r_Ir_A} \right|}.
	\end{split}
\end{align}
Then we have
\begin{align}\lb{S_gen_near_Ext}
S_{\text{gen}}=\frac{\pi r_{I}^{2}}{G}-\frac{4\pi kcr_Ir_{A}^{3}}{\left[ -t_A+r_{A}^{*}+\left( t_I-r_{I}^{*} \right) \right] \left[ t_A+r_{A}^{*}-\left( t_I+r_{I}^{*} \right) \right] \left( r_I-r_h \right) \left( r_A-r_h \right)}.
\end{align}
Extremizing eq.(\ref{S_gen_near_Ext}) over $t_I$, we obtain
\begin{align}
\frac{\partial S_{\text{gen}}}{\partial t_I}=\frac{-4\pi kcr_Ir_{A}^{3}}{\left( r_I-r_h \right) \left( r_A-r_h \right)}\frac{2\left( t_I-t_A \right)}{\left[ t_A-r_{A}^{*}-\left( t_I-r_{I}^{*} \right) \right] ^2\left[ t_A+r_{A}^{*}-\left( t_I+r_{I}^{*} \right) \right] ^2}=0,
\end{align}
which has the solution
\begin{align}\lb{t_I_Ext_near}
t_I=t_A.
\end{align}
Putting eq.(\ref{t_I_Ext_near}) into eq.(\ref{S_gen_near_Ext}), we have
\begin{align}\lb{S_gen_Ext_near}
S_{\text{gen}}=\frac{\pi r_{I}^{2}}{G}-\frac{4\pi kcr_Ir_{A}^{3}}{\left( r_{A}^{*}-r_{I}^{*} \right) ^2\left( r_I-r_h \right) \left( r_A-r_h \right)}.
\end{align}
Since both the island $\partial I$ and the cutoff surface $A$ are assumed to be  near and outside the horizon \cite{Tong:2023nvi}, we can express $r_I=r_h\left( 1+\varepsilon \right) ,\ r_A=r_h\left( 1+\sigma \right) $, where $\varepsilon < \sigma \ll 1$. Then we have
\begin{align}
r_{I}^{*}=r_I+2r_h\ln \left|\frac{ r_I-r_h}{r_h} \right| -\frac{r_{h}^{2}}{r_I-r_h}=r_h\left( 1+\varepsilon \right) +2r_h\ln \left| \varepsilon \right|-\frac{r_h}{\varepsilon},\\
r_{A}^{*}=r_A+2r_h\ln \left( \frac{r_A-r_h}{r_h} \right) -\frac{r_{h}^{2}}{r_A-r_h}=r_h\left( 1+\sigma \right) +2r_h\ln \sigma -\frac{r_h}{\sigma}.
\end{align}
Now eq.(\ref{S_gen_Ext_near}) becomes
\begin{align}\lb{S_Ext_epsilon}
	\begin{split}
S_{\text{gen}}&=\frac{\pi r_{h}^{2}\left( 1+\varepsilon \right) ^2}{G}-\frac{4\pi kcr_{h}^{4}\left( 1+\varepsilon \right) \left( 1+\sigma \right) ^3}{\left[ r_h\left( 1+\sigma \right) +2r_h\ln \sigma  -\frac{r_h}{\sigma}-\left( r_h\left( 1+\varepsilon \right) +2r_h\ln \left| \varepsilon  \right|-\frac{r_h}{\varepsilon} \right) \right] ^2\varepsilon \sigma r_{h}^{2}}\\
&\approx \frac{\pi r_{h}^{2}\left( 1+\varepsilon \right) ^2}{G}-\frac{4\pi kc\sigma \varepsilon \left( 1+\varepsilon \right) \left( 1+\sigma \right) ^3}{\left[ 2\sigma \varepsilon \ln \left( \sigma /\varepsilon \right) +\left( \sigma -\varepsilon \right) \right] ^2}.
	\end{split}
\end{align}
Extremizing eq.(\ref{S_Ext_epsilon}) over $ \varepsilon $ yields
\begin{align}\lb{PS_Pepsilon}
\frac{\partial S_{\text{gen}}}{\partial \varepsilon}=\frac{2\pi r_{h}^{2}\left( 1+\varepsilon \right)}{G}-\frac{4\pi kc\sigma \left( 1+\sigma \right) ^3\left[ \varepsilon +\sigma +4\varepsilon \sigma -2\varepsilon \sigma \ln \left( \sigma /\varepsilon \right) \right]}{\left[ -\varepsilon +\sigma +2\sigma \varepsilon \ln \left( \sigma /\varepsilon \right) \right] ^3}=0.
\end{align}
By using $\sigma \pm \varepsilon \gg 2\sigma \varepsilon \ln \left( \sigma /\varepsilon \right) $ and $\varepsilon \ll 1$ to simplify eq.(\ref{PS_Pepsilon}), we find
\begin{align}\lb{PS=0}
\frac{2\pi r_{h}^{2}}{G}-\frac{4\pi kc\sigma \left( 1+\sigma \right) ^3\left( \varepsilon +\sigma \right)}{\left( -\varepsilon +\sigma \right) ^3}=0.
\end{align}
Letting $x\equiv \frac{\varepsilon}{\sigma}(0<x<1)$ and rewriting eq.(\ref{PS=0}) lead to
\begin{align}\lb{x}
\frac{1+x}{\left( 1-x \right) ^3}=\frac{\sigma r_{h}^{2}}{2Gkc\left( 1+\sigma \right) ^3},
\end{align}
where  the right side of eq.(\ref{x}) is a constant. The function $g\left( x \right) =\frac{1+x}{\left( 1-x \right) ^3}$ monotonically increases with $x$ in the interval   $(0,1)$ and the local minimum value of function $g(x)$ is located at $x=1$ (see Fig.\ref{g_x}), that is $g(0)=1$.
Therefore, the existence of the island solution of eq.(\ref{PS=0}) requires 
\begin{align}\lb{C}
C\equiv \frac{\sigma r_{h}^{2}}{2Gkc\left( 1+\sigma \right) ^3}>1.
\end{align}
If the constraint (\ref{C}) is satisfied, there exists only one solution $x=x_0$ (i.e. $\varepsilon_0=x_0 \sigma $ ), which is the island solution. If the constraint (\ref{C}) is violated, there would be no nontrivial island solution.

We have shown the existence of the island solution for $\varepsilon <\sigma \ll 1$ in the extremal case, which requires the constraint (\ref{C}), although we do not give the exact expression of the island solution. Now we will consider the corresponding entanglement entropy of Hawking radiation under the constrain condition (\ref{C}). From eq.(\ref{C}), we find $\frac{\sigma r_{h}^{2}}{2G}>kc\left( 1+\sigma \right) ^3$. Putting $\varepsilon_0=x_0\sigma$  into eq.(\ref{S_Ext_epsilon}) yields 
\begin{align}\lb{S_C}
	\begin{split}
S_{\text{gen}}&\approx \frac{\pi r_{h}^{2}\left( 1+\varepsilon _0 \right) ^2}{G}-\frac{4\pi kc\sigma \varepsilon \left( 1+\varepsilon _0 \right) \left( 1+\sigma \right) ^3}{\left[ 2\sigma \varepsilon \ln \left( \sigma /\varepsilon _0 \right) +\left( \sigma -\varepsilon _0 \right) \right] ^2}\\
&>\frac{\pi r_{h}^{2}\left( 1+\varepsilon \right) ^2}{G}-\frac{2\pi r_{h}^{2}\sigma ^2\varepsilon _0\left( 1+\varepsilon _0 \right)}{G\left( \sigma -\varepsilon _0 \right) ^2}\\
&=\frac{\pi r_{h}^{2}}{G}\left( 1+\varepsilon _0 \right) \left[ 1+\varepsilon _0-\frac{2\sigma x_0}{\left( 1-x_0 \right) ^2} \right].
	\end{split}
\end{align}
It is evident that $\frac{x_0}{\left( 1-x_0 \right) ^2}<\frac{1+x_0}{\left( 1-x_0 \right) ^3}=C$. Using this fact,  we can derive
\begin{align}\lb{_S}
S_{\text{gen}}>\frac{\pi r_{h}^{2}}{G}\left( 1+\varepsilon _0 \right) \left( 1+\varepsilon _0-2\sigma C \right) >\frac{\pi r_{h}^{2}}{G}\left( 1-2\sigma C \right). 
\end{align}
Meanwhile,
\begin{align}\lb{S_}
	\begin{split}
S_{\text{gen}}&\approx \frac{\pi r_{h}^{2}\left( 1+\varepsilon \right) ^2}{G}-\frac{4\pi kc\sigma \varepsilon \left( 1+\varepsilon \right) \left( 1+\sigma \right) ^3}{\left( \sigma -\varepsilon \right) ^2}\\
&=\frac{\pi r_{h}^{2}}{G}-\frac{4\pi kcx_0}{\left( 1-x_0 \right) ^2}<\frac{\pi r_{h}^{2}}{G}.
	\end{split}
\end{align}
Comparing eq.(\ref{_S}) with eq.(\ref{S_}), we argue that $S_{\text{gen}}$ is approximately equal to $S_{\text{BH}}$. Thus,  the presence of an island configuration leads to a bound on the entropy of Hawking radiation at late times. This bound is determined by the black hole's Bekenstein-Hawking entropy, which decreases monotonically due to the black hole evaporation.

\subsection{Without island}

In this case, eq.(\ref{S_ent_far})
\begin{align}
S_{\text{gen}}=\frac{c}{12}\ln \left[ d\left( A,I \right)^4 e^{2\rho _A}e^{2\rho _I} \right] 
\end{align}
is also suitable, and
\begin{align}\lb{ln}
\ln \left( e^{2\rho _A}e^{2\rho _I} \right) =\ln \left[ \left( 1-\frac{r_h}{r_A} \right) ^2\left( 1-\frac{r_h}{r_I} \right) ^2\frac{\left( t_A-r_{A}^{*} \right) ^2}{4r_{h}^{2}} \right].
\end{align}
When there is no island, $r_I=0$, regardless of whether the cutoff surface is far from or near the horizon. Consequently, eq.(\ref{ln}) diverges, rendering the entropy ill-defined in these scenarios. This issue has also been observed in previous studies \cite{Kim:2021gzd, Ahn:2021chg, HosseiniMansoori:2022hok}. It is suggested in \cite{Ahn:2021chg}  that in the absence of an island, the entanglement entropy of Hawking radiation becomes divergent, indicating the breakdown of the semi-classical approximation in the vicinity of the singularity. Our viewpoint aligns with the findings of \cite{Ahn:2021chg}. As the island formula is based on semi-classical approximation, its validity is compromised in this context. Therefore, we conclude that in the extremal case, the absence of an island renders the island formula ineffective. However, it remains possible to calculate the entanglement entropy of Hawking radiation in the presence of an island, particularly in extremal cases.

\section{Conclusions}\lb{sec_con}

In this paper, we have expanded upon the method introduced in \cite{Gan:2022jay} for calculating the entanglement entropy of Hawking radiation from Schwarzschild black holes, extending it to more general charged black holes. Our approach utilizes the ``in'' vacuum state, specifically tailored for one-sided dynamical black holes, which accurately describes the vacuum in the Minkowski region. This choice distinguishes our study from previous works that relied on the Hartle-Hawking state \cite{Kim:2021gzd, Wang:2021woy},  more suitable for eternal black holes. Aligning with findings for one-sided Schwarzschild black holes, we have demonstrated the validity of the $s$-wave approximation for dynamic charged black holes.
\begin{table}[h]
		\centering 
		\begin{tabular}{|c |c| c| c| c|}
			\hline   
			RN black Hole & without island & with island& Page time \\ 
			\hline
			non-extremal case & $S_{\text{gen}}\simeq \frac{c}{12}\kappa t_A$ & $S_{\text{gen}}\simeq S_{\text{BH}}$ & $t_{Page}\simeq \frac{12}{c\kappa}S_{\text{BH}}$ \\
			\hline
			extremal case & ill-defined & $S_{\text{gen}}\simeq S_{\text{BH}}$ & ill-defined \\
			\hline
		\end{tabular}\caption{Entanglement entropy and Page time in one-sided charged black hole formed from gravitational collapse.}\label{table}
\end{table}

\begin{table}[h]
	\centering 
	\begin{tabular}{|c |c| c| c| c|}
		\hline   
		RN black Hole & without island & with island& Page time \\ 
		\hline
		non-extremal case & $S_{\text{gen}}\simeq \frac{c}{3}\kappa t_A$ & $S_{\text{gen}}\simeq 2S_{\text{BH}}$ & $t_{Page}\simeq \frac{6}{c\kappa}S_{\text{BH}}$ \\
		\hline
		extremal case & ill-defined & $S_{\text{gen}}\simeq S_{\text{BH}}$ & ill-defined \\
		\hline
	\end{tabular}\caption{Entanglement entropy and Page time in eternal charged black hole.}\label{table2}
\end{table}

\begin{figure}[h!]
    \begin{tabular}{cc}
	\includegraphics[height=6.5cm]{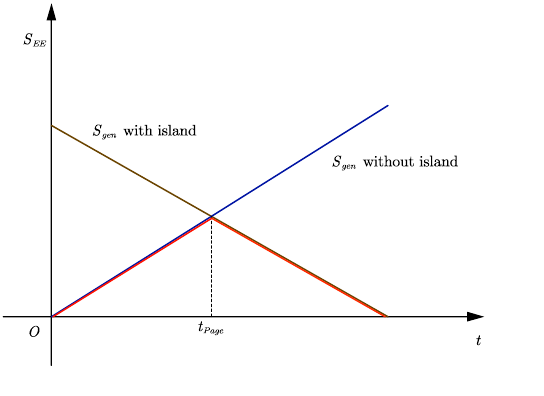}&
        \includegraphics[height=6.5cm]{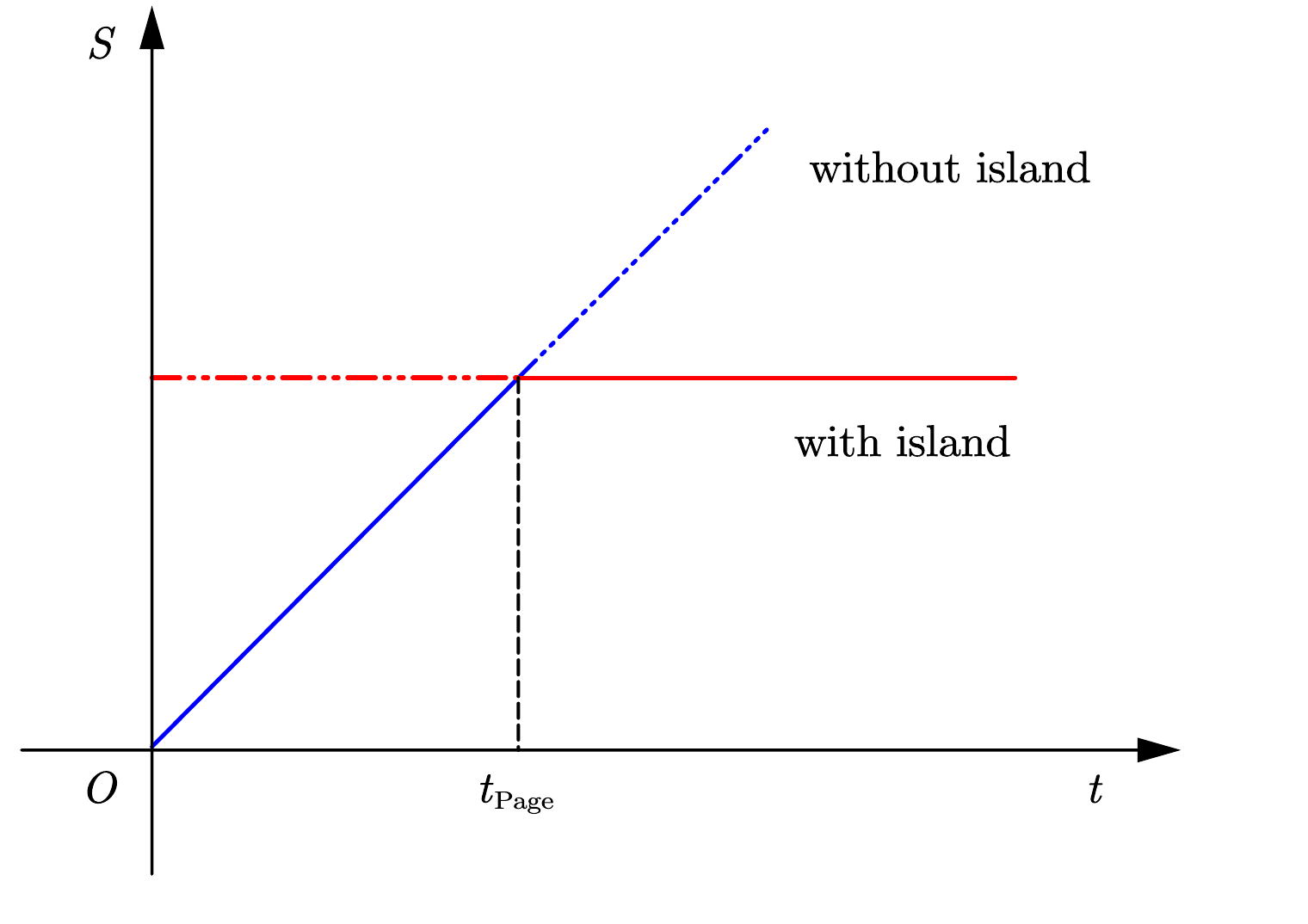}\\
	(a) & (b) \\[6pt]
    \end{tabular}
\caption{(a) Page curve (red line) for one-sided charged black hole formed by the collapse of the null shell.  (b) Page curve (solid line) for eternal charged black hole. $S_{EE}$ and $S$ represents entanglement entropy of the one sided black hole and eternal black hole, respectively. Both figures represent Page curves under non-extremal case.}\label{Page}
\end{figure}

We have separately discussed non-extremal and extremal RN black holes, as their geometric structures exhibit distinct characteristics. Our key findings are  summarized in Table \ref{table}, and the resulting Page curve of non-extermal case is illustrated in Fig.\ref{Page} (a). For non-extremal RN black holes, when the cutoff surface is far from the horizon, the island emerges at a later time, with its boundary $\partial I$ located inside and near the horizon. Conversely, when the cutoff surface is near the horizon, $\partial I$ is outside and near the horizon. In both  scenarios, the emergence of the island preserves the bound of entropy, and the entanglement entropy of Hawking radiation adheres to the Page curve. These results reduce to those for Schwarzschild black holes \cite{Gan:2022jay} when the charge is set to zero (i.e. $Q=0$). Additionally, we have explored the influence of charge on the position of $\partial I$ and discovered that for near-extremal RN black holes (where $Q$ is large), the method of calculating entanglement entropy and the position of $\partial I$ discussed in this paper are no longer applicable.

For extremal RN black holes, the behavior of entropy and the island becomes even more intriguing. We have observed that when the cutoff surface is far from the horizon, $\partial I$ is located outside and near the horizon, differing from non-extremal RN black holes. However, when the cutoff surface is near the horizon, an island emerges only  under certain constraints, with $\partial I$ located outside and near the horizon. It is important to note that in the absence of the island, the entropy becomes ill-defined. This suggests that the island formula is broken in the vicinity of the singularity.

The calculation results of the eternal charged black hole in HH vacuum state are also listed in Table \ref{table2} \cite{Tong:2023nvi}. This highlights the difference in calculation results between the ``in'' vacuum state and the HH vacuum state. After the Page time, the Page curve of the eternal charged black hole becomes a horizontal straight line (Fig.\ref{Page} (b)) \cite{Wang:2021woy,Tong:2023nvi}, which is significantly different from the Page curve we derived. It is obvious that the Page curve we obtained more clearly demonstrates the unitary evolution of a pure state.

In conclusion, our study enhances the understanding of  the black hole information paradox in charged black holes by utilizing the ``in'' vacuum state and considering both non-extremal and extremal charged black holes. The results highlight the influence of charge on the position of the island and demonstrate the preservation of entropy bounds. These findings contribute to a deeper understanding of the behavior of entanglement entropy in one-sided dynamical black hole scenarios and open avenues for further research in this field.

\section*{Acknowledgments}

H.W.Y was supported in part by the NSFC under Grant No. 12075084. F.W.S. was supported by the National Natural Science Foundation of China under the Grants No. 12375049, 11975116, and Key Program of the Natural Science Foundation of Jiangxi Province under the Grant No. 20232ACB201008. W.C.G. was supported by  the Initial Research Foundation of Jiangxi Normal University under the Grant No. 12022827.


\end{document}